\documentclass[aps,prl,twocolumn,showpacs]{revtex4}
\usepackage{amsmath,amssymb}
\usepackage{graphics,color}

\newcommand{\rv}{\vec{r}}
\newcommand{\dr}{\textrm{d}}
\newcommand{\er}{\textrm{e}}
\newcommand{\ir}{\textrm{i}}
\newcommand{\oD}{\mathsf{D}}
\newcommand{\oP}{\mathsf{P}}

\begin{document}

\title{Selective amplification of scars in a chaotic optical fiber}
\author{Claire Michel, Val\'erie Doya, Olivier Legrand, and Fabrice Mortessagne}
\email{Fabrice.Mortessagne@unice.fr}
\affiliation{Laboratoire de Physique de la Mati\`ere Condens\'ee\\
Universit\'e de Nice Sophia-Antipolis \& CNRS, UMR 6622\\ 06108 Nice cedex 2,
France}

\begin{abstract}
In this letter we propose an original mechanism to select scar modes through coherent gain amplification in a multimode D-shaped fiber. More precisely, we numerically demonstrate the selective amplification of scar modes by positioning a gain region in the vicinity of the self-focal point of the shortest periodic orbit in the transverse motion.
\end{abstract}

\pacs{05.45.Mt, 42.55.Wd}

\maketitle

Since their discovery in 1984 by Eric Heller \cite{Heller84}, scars have remained one of the most intriguing features in the field of \emph{Wave Chaos}. They constitute `anomalous' modes of wave cavities where rays exhibit chaotic dynamics. According to Random Matrix Theory \cite{Stockmannbook}, as well as from semiclassical analyses \cite{Berry77}, a spatially uniform distribution is expected for modes in such chaotic cavities. Scars deviate significantly from this ergodic behavior by exhibiting unexpected enhancement of intensity in the vicinity of the least unstable periodic orbits (PO) of the corresponding billiards. Since their numerical birth in a quantum context, scars have been experimentally observed with microwaves \cite{Microwave}, with capillary or Faraday waves \cite{Hydro}, in guided optics \cite{Doya_PRl_2002} and their technological interest is now clearly established \cite{Gensty, Microlaser, Hui}. The violation of the ergodic hypothesis implied by the existence of scars is justified by various theoretical approaches \cite{theo} but remains mathematically uncertain  \cite{math}.

In previous works we showed that a passive optical fiber with a chaotic-billiard-shape transverse section  constitutes  a powerful experimental system for studying (and imaging) wave chaos \cite{Doya_PRE_2002}, and particularly when scars are involved \cite{Doya_PRl_2002}. Indeed, at the output of a multimode D-shaped chaotic fiber we were able to put to evidence a scar mode associated to the shortest PO. Unfortunately, diffraction due to the finite aperture of the fiber precludes any truly selective excitation of a single mode. For an input illumination privileging the lowest order modes, a few tens of modes contribute to the optical field, and the best that one can do is to slightly enhance one of the lowest order scars. For larger orders, the increasing number of excited modes hampers the scar enhancement effect. Scar modes have also been observed in microlasers \cite{Microlaser}. These devices are more and more present in photonics, for which high directionality  and low lasing threshold are required. Scar modes constitute the cavity modes with the highest quality factors \cite{Hui} and privileged directions of emission. In such lasing microcavities, boundary losses play a dominant role in the mode selection mechanism and,  for a given shape, only a few efficient scar modes are supported \cite{Hui}. A different selection mechanism may be found in the domain of wave propagation in random media. In this context, the amplification of localized modes is observed with random lasers (see Ref. \cite{Cao05} for a recent review), where they play similar to scar modes. Both types of modes rely on strong coherent effects and display common statistical features \cite{Pradhan00}. Numerical investigations recently showed that, in the localized regime, the lasing modes in a random laser just above threshold are the modes of the passive disordered system \cite{Sebbah_PRB_2002}. The identification  between lasing and passive modes even holds when the gain region is smaller than the mode size\cite{Vanneste_PRL_2001}. Inspired by these results, we present an original way to achieve a selective amplification of scar modes in a highly multimode D-shaped optical fiber. More precisely, we demonstrate how scar modes can be selectively amplified by positioning a gain region in the vicinity of specific points along a short PO known to give rise to scar modes. We also illustrate the robustness of this amplification with a spatially incoherent speckle-like input.

We recently showed that multimode D-shaped fibers are perfectly suited to achieve high power optical amplification \cite{Doya_OptLett}. In the single-mode doped core of an optical amplifier, the guided optical field (called the signal) is amplified at the cost of an ancillary field (called the pump) which propagates in the multimode D-shaped silica cladding \cite{Desurvire}. Usually, the optical index of the Erbium or Ytterbium doped core  is higher than the index of the surrounding D-shaped silica structure. Therefore, the signal is guided along the doped region, and, by guiding the pump, the multimode external core only plays the role of an energy reservoir. By some appropriate modifications of the fabrication process, which will be presented in a forthcoming publication, it is possible to obtain a negligible index mismatch between the small active region and the rest of the fiber,  at least sufficiently small to ensure that no mode is guided inside the doped area. Thus, both the signal and the pump propagate in the entire section of the multimode D-shaped fiber. Here we propose a realistic numerical simulation of such a chaotic fiber with a localized Ytterbium-doped region. Ytterbium in a silica matrix may be viewed as a four-level system with two metastable levels, which, in our case, can be treated as an effective two-level system. The wavelength is $\lambda_s=1020$\,nm, for the signal, and $\lambda_p=980\,$nm, for the pump, thus limiting the signal reabsorption. The simulation is based on the \emph{Beam Propagation Method} (BPM) \cite{Feit}, which we already successfully used for simulating optical amplification \cite{Doya_OptLett}. 

\begin{figure}[h]
\begin{center}
\begin{picture}(0,0)%
\includegraphics{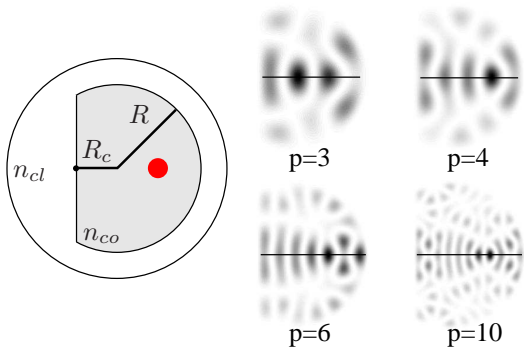}%
\end{picture}%
\setlength{\unitlength}{4144sp}%
\begingroup\makeatletter\ifx\SetFigFontNFSS\undefined%
\gdef\SetFigFontNFSS#1#2#3#4#5{%
  \reset@font\fontsize{#1}{#2pt}%
  \fontfamily{#3}\fontseries{#4}\fontshape{#5}%
  \selectfont}%
\fi\endgroup%
\begin{picture}(3108,2085)(-16,-1261)
\put(449,-595){\makebox(0,0)[lb]{\smash{{\SetFigFontNFSS{10}{12.0}{\rmdefault}{\mddefault}{\updefault}{$n_{co}$}}}}}
\put(436,-109){\makebox(0,0)[lb]{\smash{{\SetFigFontNFSS{10}{12.0}{\familydefault}{\mddefault}{\updefault}{$R_c$}}}}}
\put( 29,-220){\makebox(0,0)[lb]{\smash{{\SetFigFontNFSS{10}{12.0}{\rmdefault}{\mddefault}{\updefault}{$n_{cl}$}}}}}
\put(721,101){\makebox(0,0)[lb]{\smash{{\SetFigFontNFSS{10}{12.0}{\familydefault}{\mddefault}{\updefault}{$R$}}}}}
\end{picture}
\end{center}
\caption {Transverse cross section of the D-shaped multimode fiber and 4 scar modes associated to the shortest periodic orbit.}
\label{figfibre}
\end{figure}

The optical fiber we consider has the D-shaped cross section shown in figure \ref{figfibre}. Its highly multimode core is made of pure silica of refractive index $n_{co}=1.451$ and has a $2R=125\,\mu$m diameter. The value of the truncated radius is $R_c=R/2$. This core is surrounded by a polymer cladding of refractive index  $n_{cl}=1.41$ that acts both as an optical cladding (to guide light in the core) and as a mechanical cladding (to protect the core). By using Dirichlet boundary conditions (metallic fiber), we are able to solve numerically the Helmholtz equation and we show on Fig. \ref{figfibre} some scar modes associated to the 2-bounce periodic orbit (2PO). As the losses of our system are weak, guided modes of the metallic and dielectric fibers are quantitatively the same for the first hundreds of modes. The order $p$ of a given scar gives  the corresponding transverse wavenumber $\kappa^{(p)}$ (eigenvalue of Eq. (\ref{statschro}) below) through the following Bohr-Sommerfeld-like quantization relation:
\begin{equation}
\kappa^{(p)}\mathcal{L}-\Delta\phi-\frac{\pi}{2}=2\pi p
\end{equation}
where $\mathcal{L}$ is the length of the 2PO and $\Delta\phi$ is the phaseshift at the core/cladding interface. The additional $\pi/2$ phaseshift is due to the unique self-focal point of the 2PO. The active Ytterbium ions are located in a disk of diameter 15\,$\mu$m enclosing this self-focal point (see Fig. \ref{figfibre}), that also corresponds to a maximum of intensity of the vast majority of the 2PO scars. 

We denote by $z$ the position along the axis of the fiber and by $\rv$ the position in the transverse plane. We showed in \cite{Doya_PRl_2002} that the scalar approximation is legitimate in the weak guidance limit. Thus the electromagnetic field $\psi_s(\rv,z)$ representing the signal obeys, in a passive fiber, the scalar three-dimensional Helmholtz stationary equation
\begin{equation}
\label{helm3d}
\left(\Delta_{\bot}+\partial_{zz}\right)\psi_s(\rv,z)
+n^2(\rv)k_0^2\psi_s(\rv,z)=0
\end{equation}
where $\Delta_{\bot}$ is the transverse Laplacian and $k_0=2\pi/\lambda_s$ is the vacuum wavenumber. Using the translational invariance of the refractive index, Eq. (\ref{helm3d}) can be reduced to a stationary Schr\"odinger equation 
\begin{equation}
\label{statschro}
\left[-\frac{1}{2}\Delta_{\bot}+V(\rv)\right]\phi(\rv;\beta)=\frac{1}{2}\kappa^2\phi(\rv;\beta)
\end{equation}
where $\phi$ is given by $\psi_s(\rv,z)=\int\dr\beta\phi(\rv;\beta)\er^{\ir\,\beta z}$, the  potential $V(\rv)=(\beta_{co}^2-n^2(\rv)k_0^2)/2$ with $\beta_{co} \equiv
n_{co}k_0$ and the transverse wavenumber  $\kappa^2=\beta_{co}^2-\beta^2$ which takes on discrete values $\kappa_n$ associated to guided modes $\phi_n$. Any guided solution $\psi_s(\rv,z)$ of Eq. (\ref{helm3d}) can be decomposed on the basis generated by the bound eigenstates \(\phi_n(\rv)=\phi(\rv;\beta_n)\) of the Schr\"odinger equation (\ref{statschro}): 
\begin{equation}
\psi_s(\rv,z)=\sum_nc_n\phi_n(\rv)\exp\left(\ir\,\beta_nz\right) \, .
\end{equation}
To each mode can be associated an angle $\theta_n$ with respect to the $z$-axis defined by $\sin\theta_n=\kappa_n/\beta_{co}$. The cut-off angle for guided modes is given by $\sin\theta_{max}=\sqrt{1-(n_{cl}/n_{co})^2}\simeq 0.34$, which corresponds to the maximum value $\kappa_{max}=\sqrt{\beta_{co}^2-\beta_{cl}^2}$. Knowing this value, one can obtain the total number of guided modes at wavelength $\lambda_s$ which is approximately $3\,500$ \cite{Doya_PRE_2002}. \textit{Mutatis mutandis}, we can derive similar expressions for the electromagnetic field $\psi_p(\rv,z)$ of the pump. 

In the paraxial approximation (see \cite{Doya_PRE_2002} for a thorough  discussion of this approximation in the present context), $z$ plays the role of time in a Schr\"odinger-like equation. By using the standard BPM numerical scheme, the coupled evolution of the fields over an infinitesimal pseudo-time step may be written
\begin{multline}
\psi_{s,p}(\rv,z+\dr z)=\exp\left[-\ir(\oD_{s,p}+\oP_{s,p})\dr z\right.\\ +\left.\frac{1}{2}\alpha_{s,p}(\rv,z)\dr z\right]\psi_{s,p}(\rv,z)
\label{BPM}
\end{multline}
where $\oD=-(1/2\beta_{co})\Delta_{\bot}$ is the `diffraction operator' acting in the Fourier space and $\oP=-(1/2\beta_{co})(n^2(\rv)k_0^2-\beta_{co}^2)$ is the `propagating operator' accounting for the guided propagation in the real space. From the population densities $N_1(\rv,z)$ and $N_2(\rv,z)$ of the two effective Ytterbium energy levels, one obtains the two amplification factors: $\alpha_p(\rv,z)=-\sigma_{pa}N_1(\rv,z)$ and $\alpha_s(\rv,z)=\sigma_{sa}\big(\eta_sN_2(\rv,z)-N_1(\rv,z)\big)$, where $\sigma_{sa}$ ($\sigma_{pa}$) is the absorption cross section of the signal (pump), the stimulated emission cross section at the wavelength of the signal $\sigma_{se}$ is also involved through the ratio $\eta_s=\sigma_{se}/\sigma_{sa}$ (for a complete account see Ref. \cite{THY}, section 14.4). In all the results presented below $\alpha_s\lesssim 0.1\,\textrm{m}^{-1}$, preventing any significant gain-guiding effect. This effect has been recently proposed as a potentially effective mechanism for providing single-mode operation of high-power fiber lasers \cite{Siegman}. 

In all the simulations performed, the numerical initial condition used for the pump evolution (\ref{BPM}) corresponds to an illumination by a focussed laser beam, ensuring an optimal coupling with a large number of generically ergodic modes. The input powers are 6\,W for the pump and 100\,$\mu$W for the signal (corresponding to the so-called weak signal regime \cite{Desurvire}), and the cross sections have the following values: $\sigma_{pa}=2.65\times 10^{-24}\,$m$^2$, $\sigma_{sa}=5.56\times 10^{-26}\,$m$^2$, $\sigma_{se}=6.00\times 10^{-25}\,$m$^2$. The identification of the guided modes which are present in the propagating signal is performed through what we call the pseudo-time-frequency spectrum $C(\kappa;z)$. This quantity is derived from a standard procedure in quantum physics adapted to the numerical algorithm we use \cite{FEI}. More precisely, $C(\kappa;z)$  is obtained from the Fourier transform of the correlation function defined as the overlap of the propagating field $\psi_s(\rv,z)$ with the initial condition along a propagation length $L_z$:
\begin{eqnarray}
C(\kappa;z)&=&\int\limits_{z-\frac{L_z}{2}}^{z+\frac{L_z}{2}}\dr z'\iint\dr \rv\,\psi_s^*(\rv,0)\psi_s(\rv,z')\er^{-\ir\beta(\kappa) z'}\nonumber\\
&\sim&\sum_nA_n(z)\delta_\epsilon(\kappa_n-\kappa)
\label{Ckt}
\end{eqnarray}
where, for the sake of simplicity, we have written $C(\kappa;z)$ in a form which clearly  exhibits smoothed delta peaks at the values of the transverse wavenumber associated to the eigenmodes of the fiber.

\begin{figure}[h]
\begin{center}
\begin{picture}(0,0)%
\includegraphics{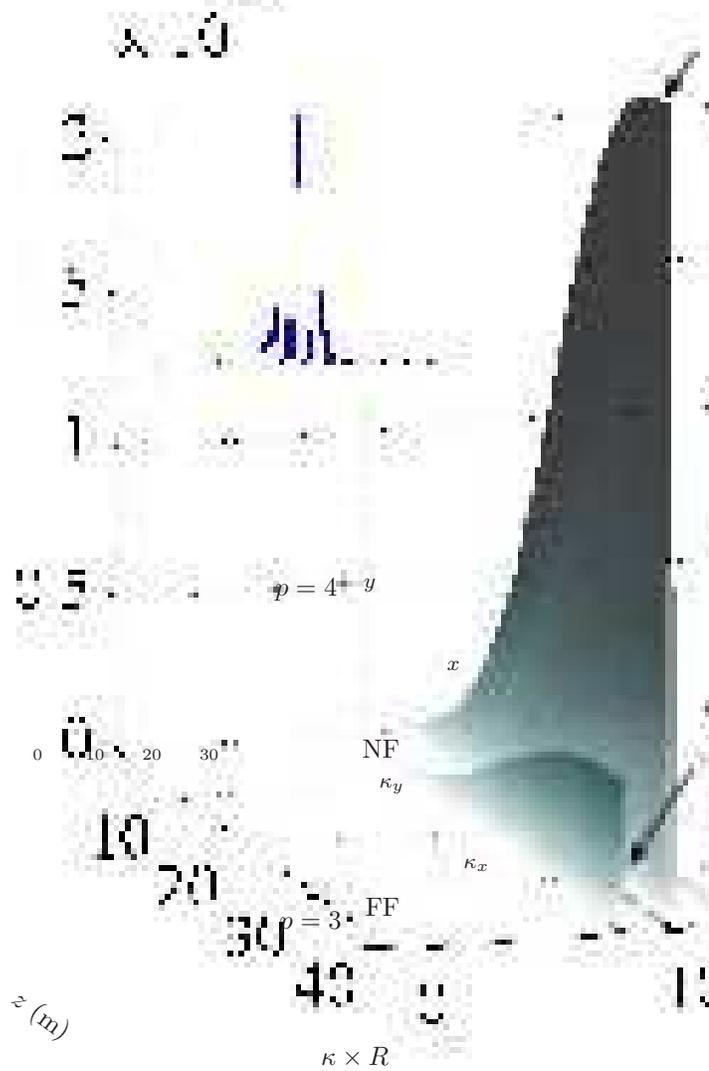}%
\end{picture}%
\setlength{\unitlength}{4144sp}%
\begingroup\makeatletter\ifx\SetFigFontNFSS\undefined%
\gdef\SetFigFontNFSS#1#2#3#4#5{%
  \reset@font\fontsize{#1}{#2pt}%
  \fontfamily{#3}\fontseries{#4}\fontshape{#5}%
  \selectfont}%
\fi\endgroup%
\begin{picture}(4127,3106)(-3,-2279)
\put(3213,-1041){\makebox(0,0)[lb]{\smash{{\SetFigFontNFSS{8}{9.6}{\rmdefault}{\mddefault}{\updefault}{$\kappa_x$}}}}}
\put(3113,146){\makebox(0,0)[lb]{\smash{{\SetFigFontNFSS{8}{9.6}{\rmdefault}{\mddefault}{\updefault}{$x$}}}}}
\put(2615,638){\makebox(0,0)[lb]{\smash{{\SetFigFontNFSS{8}{9.6}{\rmdefault}{\mddefault}{\updefault}{$y$}}}}}
\put(2112,-1404){\makebox(0,0)[lb]{\smash{{\SetFigFontNFSS{10}{12.0}{\rmdefault}{\mddefault}{\updefault}{$p=3$}}}}}
\put(2086,594){\makebox(0,0)[lb]{\smash{{\SetFigFontNFSS{10}{12.0}{\rmdefault}{\mddefault}{\updefault}{$p=4$}}}}}
\put(2607,-378){\makebox(0,0)[lb]{\smash{{\SetFigFontNFSS{10}{12.0}{\rmdefault}{\mddefault}{\updefault}{NF}}}}}
\put(2624,-1326){\makebox(0,0)[lb]{\smash{{\SetFigFontNFSS{10}{12.0}{\rmdefault}{\mddefault}{\updefault}{FF}}}}}
\put(2709,-558){\makebox(0,0)[lb]{\smash{{\SetFigFontNFSS{8}{9.6}{\rmdefault}{\mddefault}{\updefault}{$\kappa_y$}}}}}
\put(500,-1844){\rotatebox{324.0}{\makebox(0,0)[lb]{\smash{{\SetFigFontNFSS{10}{12.0}{\rmdefault}{\mddefault}{\updefault}{$z$ (m)}}}}}}
\put(2359,-2215){\makebox(0,0)[lb]{\smash{{\SetFigFontNFSS{10}{12.0}{\rmdefault}{\mddefault}{\updefault}{$\kappa\times R$}}}}}
\put(636,-389){\makebox(0,0)[lb]{\smash{{\SetFigFontNFSS{6}{7.2}{\rmdefault}{\mddefault}{\updefault}{0}}}}}
\put(1631,-389){\makebox(0,0)[lb]{\smash{{\SetFigFontNFSS{6}{7.2}{\rmdefault}{\mddefault}{\updefault}{30}}}}}
\put(1289,-389){\makebox(0,0)[lb]{\smash{{\SetFigFontNFSS{6}{7.2}{\rmdefault}{\mddefault}{\updefault}{20}}}}}
\put(947,-389){\makebox(0,0)[lb]{\smash{{\SetFigFontNFSS{6}{7.2}{\rmdefault}{\mddefault}{\updefault}{10}}}}}
\end{picture}%
\end{center}
\caption {Two-dimensional plot of  $C(\kappa;z)$ defined in (\ref{Ckt}) for $L_z=1.3\,$m. Left inset: spectrum at $z=0$. Right Inset: near-field (NF) and far-field (FF) intensities at the maximum signal amplification length $z_{max}=39\,$m.}
\label{PlaneWave}
\end{figure}

First, we use a selective illumination condition along the 2PO scar mode of order $p=4$ which exhibits a good overlap with the gain region. Practically, the input signal is a plane wave with the transverse wavenumber $\kappa^{(4)}=10.94/R$. The left inset in Fig. \ref{PlaneWave} shows the spectrum at $z=0$. In spite of the plane wave illumination, whose wavenumber is associated to the highest peak, the unavoidable diffraction generates extra wavenumbers in the spectrum. In a passive fiber this spectrum would be unchanged during propagation. In Fig. \ref{PlaneWave}, we show the evolution of $C(\kappa;z)$ computed with $L_z=1.3\,$m and for $z$ varying by steps of $0.65\,$m. The $\kappa^{(4)}$-mode is clearly enhanced during the propagation. But, as a consequence of the initial diffraction, the scar of order $p=3$ is also excited and amplified. It is worth noting that the spectral content of the signal is deeply modified by the amplification: two scar modes completely dominate the spectrum at length $z_{max}=39\,$m where the $\kappa^{(4)}$-peak reaches a plateau. This length corresponds to the maximum amplification length after which the mode is depleted due to the absorption of the signal by the active medium. At this optimal length, the domination of the scar of order $p=4$ can be directly visualized by imaging either the near-field intensity (NF) at the output of the fiber or the far-field intensity (FF) obtained in the focal plane of a lens. This is shown in right inset of Fig. \ref{PlaneWave}. The FF intensity exhibits two symmetric spots located  at $\pm\kappa^{(4)}$ on the $\kappa_x$ axis. Moreover, NF and FF pictures are quantitatively close to those numerically obtained for the $p=4$ scar mode of a metallic fiber (see Fig. \ref{figfibre}). We verified that none of the scar modes, even the dominant $\kappa^{(4)}$, are preferentially amplified when the doped area is shifted off the 2PO axis. These results clearly demonstrate that a properly positioned gain area can selectively amplify a single scar mode. 

\begin{figure}[h]
\begin{picture}(0,0)%
\includegraphics{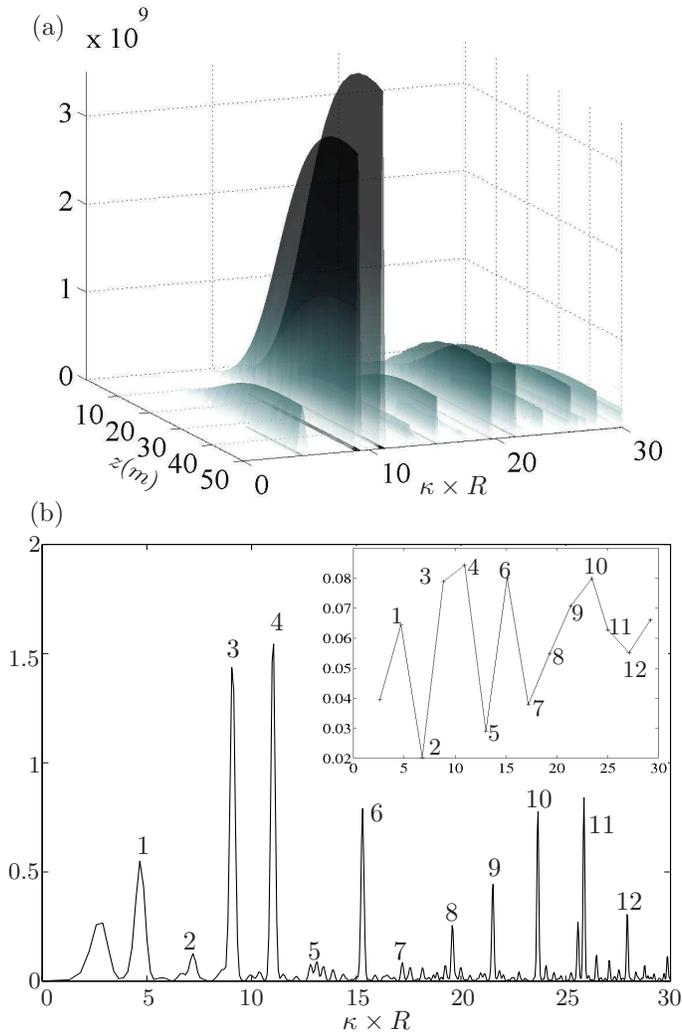}%
\end{picture}%
\setlength{\unitlength}{4144sp}%
\begingroup\makeatletter\ifx\SetFigFontNFSS\undefined%
\gdef\SetFigFontNFSS#1#2#3#4#5{%
  \reset@font\fontsize{#1}{#2pt}%
  \fontfamily{#3}\fontseries{#4}\fontshape{#5}%
  \selectfont}%
\fi\endgroup%
\begin{picture}(4220,6224)(-39,-5397)
\put(183,-5089){\makebox(0,0)[lb]{\smash{{\SetFigFontNFSS{9}{10.8}{\rmdefault}{\mddefault}{\updefault}{0}}}}}
\put( 63,-4432){\makebox(0,0)[lb]{\smash{{\SetFigFontNFSS{9}{10.8}{\rmdefault}{\mddefault}{\updefault}{0.5}}}}}
\put(183,-3794){\makebox(0,0)[lb]{\smash{{\SetFigFontNFSS{9}{10.8}{\rmdefault}{\mddefault}{\updefault}{1}}}}}
\put( 63,-3157){\makebox(0,0)[lb]{\smash{{\SetFigFontNFSS{9}{10.8}{\rmdefault}{\mddefault}{\updefault}{1.5}}}}}
\put(183,-2471){\makebox(0,0)[lb]{\smash{{\SetFigFontNFSS{9}{10.8}{\rmdefault}{\mddefault}{\updefault}{2}}}}}
\put(259,-5186){\makebox(0,0)[lb]{\smash{{\SetFigFontNFSS{9}{10.8}{\rmdefault}{\mddefault}{\updefault}{0}}}}}
\put(867,-5186){\makebox(0,0)[lb]{\smash{{\SetFigFontNFSS{9}{10.8}{\rmdefault}{\mddefault}{\updefault}{5}}}}}
\put(1445,-5186){\makebox(0,0)[lb]{\smash{{\SetFigFontNFSS{9}{10.8}{\rmdefault}{\mddefault}{\updefault}{10}}}}}
\put(2083,-5186){\makebox(0,0)[lb]{\smash{{\SetFigFontNFSS{9}{10.8}{\rmdefault}{\mddefault}{\updefault}{15}}}}}
\put(2710,-5186){\makebox(0,0)[lb]{\smash{{\SetFigFontNFSS{9}{10.8}{\rmdefault}{\mddefault}{\updefault}{20}}}}}
\put(3337,-5186){\makebox(0,0)[lb]{\smash{{\SetFigFontNFSS{9}{10.8}{\rmdefault}{\mddefault}{\updefault}{25}}}}}
\put(3946,-5186){\makebox(0,0)[lb]{\smash{{\SetFigFontNFSS{9}{10.8}{\rmdefault}{\mddefault}{\updefault}{30}}}}}
\put(202,623){\makebox(0,0)[lb]{\smash{{\SetFigFontNFSS{10}{12.0}{\rmdefault}{\mddefault}{\updefault}{(a)}}}}}
\put(187,-2287){\makebox(0,0)[lb]{\smash{{\SetFigFontNFSS{10}{12.0}{\rmdefault}{\mddefault}{\updefault}{(b)}}}}}
\put(1109,-4855){\makebox(0,0)[lb]{\smash{{\SetFigFontNFSS{10}{12.0}{\rmdefault}{\mddefault}{\updefault}{$2$}}}}}
\put(3157,-4001){\makebox(0,0)[lb]{\smash{{\SetFigFontNFSS{10}{12.0}{\rmdefault}{\mddefault}{\updefault}{$10$}}}}}
\put(1854,-4914){\makebox(0,0)[lb]{\smash{{\SetFigFontNFSS{10}{12.0}{\rmdefault}{\mddefault}{\updefault}{$5$}}}}}
\put(2367,-4914){\makebox(0,0)[lb]{\smash{{\SetFigFontNFSS{10}{12.0}{\rmdefault}{\mddefault}{\updefault}{$7$}}}}}
\put(2677,-4682){\makebox(0,0)[lb]{\smash{{\SetFigFontNFSS{10}{12.0}{\rmdefault}{\mddefault}{\updefault}{$8$}}}}}
\put(2932,-4413){\makebox(0,0)[lb]{\smash{{\SetFigFontNFSS{10}{12.0}{\rmdefault}{\mddefault}{\updefault}{$9$}}}}}
\put(3698,-4618){\makebox(0,0)[lb]{\smash{{\SetFigFontNFSS{10}{12.0}{\rmdefault}{\mddefault}{\updefault}{$12$}}}}}
\put(831,-4281){\makebox(0,0)[lb]{\smash{{\SetFigFontNFSS{10}{12.0}{\rmdefault}{\mddefault}{\updefault}{$1$}}}}}
\put(1368,-3102){\makebox(0,0)[lb]{\smash{{\SetFigFontNFSS{10}{12.0}{\rmdefault}{\mddefault}{\updefault}{$3$}}}}}
\put(1632,-2942){\makebox(0,0)[lb]{\smash{{\SetFigFontNFSS{10}{12.0}{\rmdefault}{\mddefault}{\updefault}{$4$}}}}}
\put(2522,-2663){\makebox(0,0)[lb]{\smash{{\SetFigFontNFSS{9}{10.8}{\rmdefault}{\mddefault}{\updefault}{$3$}}}}}
\put(2808,-2610){\makebox(0,0)[lb]{\smash{{\SetFigFontNFSS{9}{10.8}{\rmdefault}{\mddefault}{\updefault}{$4$}}}}}
\put(3504,-2604){\makebox(0,0)[lb]{\smash{{\SetFigFontNFSS{9}{10.8}{\rmdefault}{\mddefault}{\updefault}{$10$}}}}}
\put(3432,-2884){\makebox(0,0)[lb]{\smash{{\SetFigFontNFSS{9}{10.8}{\rmdefault}{\mddefault}{\updefault}{$9$}}}}}
\put(3653,-2967){\makebox(0,0)[lb]{\smash{{\SetFigFontNFSS{9}{10.8}{\rmdefault}{\mddefault}{\updefault}{$11$}}}}}
\put(2998,-2629){\makebox(0,0)[lb]{\smash{{\SetFigFontNFSS{9}{10.8}{\rmdefault}{\mddefault}{\updefault}{$6$}}}}}
\put(2521,-2131){\makebox(0,0)[lb]{\smash{{\SetFigFontNFSS{10}{12.0}{\rmdefault}{\mddefault}{\updefault}{$\kappa\times R$}%
}}}}
\put(2352,-2901){\makebox(0,0)[lb]{\smash{{\SetFigFontNFSS{9}{10.8}{\rmdefault}{\mddefault}{\updefault}{$1$}}}}}
\put(2933,-3605){\makebox(0,0)[lb]{\smash{{\SetFigFontNFSS{9}{10.8}{\rmdefault}{\mddefault}{\updefault}{$5$}}}}}
\put(3199,-3455){\makebox(0,0)[lb]{\smash{{\SetFigFontNFSS{9}{10.8}{\rmdefault}{\mddefault}{\updefault}{$7$}}}}}
\put(3318,-3146){\makebox(0,0)[lb]{\smash{{\SetFigFontNFSS{9}{10.8}{\rmdefault}{\mddefault}{\updefault}{$8$}}}}}
\put(3738,-3223){\makebox(0,0)[lb]{\smash{{\SetFigFontNFSS{9}{10.8}{\rmdefault}{\mddefault}{\updefault}{$12$}}}}}
\put(2063,-5333){\makebox(0,0)[lb]{\smash{{\SetFigFontNFSS{10}{12.0}{\rmdefault}{\mddefault}{\updefault}{$\kappa\times R$}}}}}
\put(3534,-4157){\makebox(0,0)[lb]{\smash{{\SetFigFontNFSS{10}{12.0}{\rmdefault}{\mddefault}{\updefault}{$11$}}}}}
\put(2579,-3690){\makebox(0,0)[lb]{\smash{{\SetFigFontNFSS{9}{10.8}{\rmdefault}{\mddefault}{\updefault}{$2$}}}}}
\put(2228,-4087){\makebox(0,0)[lb]{\smash{{\SetFigFontNFSS{10}{12.0}{\rmdefault}{\mddefault}{\updefault}{$6$}}}}}
\put(659,-1952){\rotatebox{331.0}{\makebox(0,0)[lb]{\smash{{\SetFigFontNFSS{9}{10.8}{\rmdefault}{\mddefault}{\updefault}{$z (m)$}}}}}}
\end{picture}%
\caption {(a) Evolution of $C(\kappa;z)$ along the fiber for an incoherent initial illumination. (b) Instantaneous spectrum averaged over 5 different incoherent illuminations. The inset shows the overlap integral between the intensity of the calculated scar modes and the doped area.}
\label{IncoherentWave}
\end{figure}

We now consider for the signal a spatially incoherent illumination resulting from a random superposition of a large number of plane waves.  We computed the coupled evolutions (\ref{BPM}) for such a random initial condition with transverse wave numbers up to $\tilde{\kappa}=\kappa_{max}/4$. The corresponding initial signal spectrum was thus almost uniform from 0 to $\tilde{\kappa}$. As depicted in Fig. \ref{IncoherentWave} (a), the evolution of $C(\kappa;z)$ exhibits a drastic mode selection along propagation. Each noticeable peak is associated to a scarred transverse wavenumber. This is emphasized in Fig. \ref{IncoherentWave} (b) where $C(\kappa;z_{max})$ is plotted for $z_{max}=45\,$m and where an averaging over five different incoherent illuminations has been performed. Almost all scar modes up to order $p=12$ ($\kappa^{(12)} R=27.1$) are significantly amplified. The first peak corresponds to light propagating into the fundamental guided mode. The differential amplification of the scar modes can be directly related to the relative intensities of the modes on the gain zone. This is illustrated in the inset of Fig. \ref{IncoherentWave} (b) where the overlap integral between the doped area and the intensity of the 2PO scar modes in a metallic fiber has been plotted. The modes of order $p=1,3,4,6$ and 10 have a good overlap with the gain region and naturally arise in the averaged pseudo-intensity spectrum. On the contrary, modes that present a bad overlap as those associated to order $p=2, 5$ and 7 are clearly less amplified. 

In this letter we addressed the problem of scar mode selection through coherent gain amplification. More specifically, in a multimode D-shaped optical fiber we demonstrated the selective amplification of scar modes by positioning a gain region in the vicinity of the self-focal point of the shortest periodic orbit in the transverse motion. This selection mechanism is so efficient that it can extract a family of scar modes from a spatially incoherent input signal. Our results may be viewed as the first step toward nonlinear wave chaos in guided optics. Indeed, pure scar modes could be optically injected in media presenting cubic nonlinearities giving rise to Kerr, four-wave mixing or Brillouin effects. Due to its non-uniform spatial intensity distribution a scar mode would not be homogeneously affected by nonlinear dephasing, giving rise to an interesting competition between the orderly scarring phenomenon and destabilizing third-order nonlinear effects.

\begin{acknowledgments}
We gratefully acknowledge fruitful discussions with P. Aschi\'eri and W. Blanc. We also wish to thank M. Ud\'e and S. Trzesien for providing physical data concerning the fiber amplifier. This work is supported by ANR contract No. 05-JCJC-0099-01.
\end{acknowledgments}


\begin{thebibliography}{}
\bibitem{Heller84} E. J. Heller, \prl {\bf 53}, 1515 (1984).

\bibitem{Microwave} A. Kudrolli, V. Kidambi, and S. Sridhar, \prl {\bf 75}, 822 (1995); J. Stein and H.-J. St\"ockmann, \textit{ibid.} {\bf 68}, 2867 (1992); S. Sridhar and E. J. Heller, Phys. Rev. A {\bf 46}, 1728 (1992).

\bibitem{Hydro} R. Bl\"umel \textit{et al.}, \pra {\bf 45}, 2641--2644 (1992); A. Kudrolli, M. C. Abraham, and J. P. Gollub,\pre {\bf 63}, 026208 (2001).

\bibitem{Doya_PRl_2002} V. Doya \textit{et al.}, \prl {\bf 88}, 014102 (2002).

\bibitem{Gensty} T. Gensty \textit{et al.}, \prl \textbf{94}, 233901 (2005).

\bibitem{Microlaser} C. Gmachl \textit{et al.}, Science {\bf 280}, 1556 (1998); C. Gmachl \textit{et al.}, Opt. Lett. {\bf 27}, 824 (2002); T. Harayama \textit{et al.}, \pre {\bf 67}, 015207 (2003).

\bibitem{Hui} W. Fang, A. Yamilov, and H. Cao, \pra {\bf 72}, 023815 (2005); W. Fang, H. Cao, and G. S. Salomon, \apl \textbf{90}, 081108 (2007).

\bibitem{Stockmannbook} H.-J. St\"ockmann, \textit{Quantum Chaos: an introduction},
Cambridge University Press (1999).

\bibitem{Berry77} M. V. Berry, J. Phys. A {\bf 10}, 2083 (1977).

\bibitem{theo} E.B. Bogomolny, Physica D  {\bf 31}, 169  (1988); M.V. Berry, Proc R. Soc. Lond. Ser. A {\bf 423}, 219 (1989); O. Agam, S. Fishman, \prl {\bf 73}, 806 (1994); L. Kaplan, E.J. Heller, \pre {\bf 59}, 6609 (1999). 

\bibitem{math} E. Lindenstrauss, Ann. of Math. {\bf 163}, 165 (2006); F. Faure and S. Nonnenmacher, Commun. Math. Phys. {\bf 245}, 201 (2004).

\bibitem{Doya_PRE_2002} V. Doya \textit{et al.}, \pre {\bf 65}, 056223 (2002).

\bibitem{Cao05} H. Cao, J. Phys. A: Math. Gen. {\bf 38}, 10497 (2005).

\bibitem{Pradhan00} P. Pradhan and S. Sridhar, \prl {\bf 85}, 2360 (2000).

\bibitem{Sebbah_PRB_2002} P. Sebbah and C. Vanneste, \prb  {\bf 66}, 
144202 (2002).

\bibitem{Vanneste_PRL_2001} C. Vanneste and P. Sebbah, \prl  {\bf 87}, 
183903 (2001).

\bibitem{Doya_OptLett} V. Doya, O. Legrand, and F. Mortessagne, Opt. Lett. {\textbf 26}, 872 (2001). 


\bibitem{Desurvire} E. Desurvire, \textit{Erbium doped fiber amplifiers}, Wiley interscience, New York (1994).


\bibitem{Feit} M. D. Feit and J. A. Fleck, Appl. Opt., {\textbf 17}, 3990 (1978).

\bibitem{THY} A. Ghatak, K. Thyagarajan, \textit{Introduction to fiber optics}, Cambridge university press (1998).

\bibitem{FEI} M. D. Feit, J. A. Fleck, Appl. Opt., \textbf{19}, 1154 (1980).

\bibitem{Siegman} A. E. Siegman, J. Opt. Soc. Am. A, \textbf{20}, 1617 (2003).

\end{thebibliography}
\end{document}